# Leveraging protection and efficiency of query answering in heterogenous RDF data using blockchain

Sara Hosseinzadeh Kassani, Kevin A. Schneider, Ralph Deters

Department of Computer Science, University of Saskatchewan, Canada

sara.kassani@mail.usask.ca, kevin.schneider@usask.ca, deters@cs.usask.ca

*Abstract*— In recent years, the digital world has experienced a massive amount of data being captured in various domain due to improvements in technology. Accordingly, big data management has emerged for storing, managing and extracting valuable knowledge from collected data. Due to the explosion of the amount of data, developing tools for accurate and timely integration of inherently heterogeneous data has become a critical need. In the first part of this study, we focus on a semantic data integration approach with a case study of a plant ontology to provide a uniform query interface between users and different data sources. In the second part of this study, we propose a distributed Hyperledger-based architecture to ensure data security and privacy preservation in a semantic data integration framework. Data privacy and security can potentially be violated by unauthorized users, or malicious entities. The proposed view layer architecture between heterogeneous data sources and user interface layer using distributed Hyperledger can ensure only authorized users have access to the data sources in order to protect the system against unauthorized violation and determine the degree of users' permission for read and write access.

## 1. Introduction

The World Wide Web (WWW), invented by Tim Berners Lee, is considered as one of the main sources for accessing information and is based primarily on Hypertext Markup Language (HTML), Uniform Resource Locator (URL), and Hypertext Transfer Protocol (HTTP) [1]. Data on the world wide web is available in different formats such as structured, semi-structured, and unstructured from various resources in a decentralized and dynamic environment [2][3]. The idea behind Semantic Web is about integration and combination of data drawn from diverse sources to have a distributed web data. URIs (Uniform Resource Identifiers) allows data items to be uniquely identified and avoids unnecessary ambiguities when referring to resources while providing semantics [4][5]. The idea of a semantic web, unlike traditional web, makes data not only human understandable but also machine interpretable [6][7]. This approach allows sharing and connecting data on the web from different sources, and is called linked data [8].

RDF is the World Wide Web Consortium (W3C) Standard language for representing information on the Web. It provides a flexible data model that facilitate description and exchange of the semantics of web content. Data in RDF is composed of three parts, namely subject, object and predicate statements, which is called a triplet. In this manner, RDF provides the tools to construct data integration in Semantic Web. SPARQL Protocol And RDF Query Language (SPARQL) is a W3C standard language for executing complex queries over RDF graphs [9][10].

The outline of this paper is as follows: Section 2 addresses the main challenges to current semantic data integration systems. A detailed description of the study design and experiment setup for designing and implementing an ontology for Arabidopsis Thaliana and ontology evaluation is given in Section 3. In Section 4, we provide a detailed discussion of the proposed architecture of view layer using distributed Hyperledger technology. Finally, Section 5, draws the conclusions.

## 2. Challenges in semantic data integration

### 2.1. Addressing data security issues

In the current information age, with the increasing number of available data from various sources such as technology, education, healthcare, insurance, finance/banking [11][12], big data management plays an important role in scientific exploration and knowledge discovery through effective collaboration among researchers [13][14]. Although scalable big data storage for analyzing and querying large amounts of data exist today, they are not fully capable to adequately meet the needs of collaboration among participants [15].

To overcome the aforementioned challenges, data integration becomes important when building an infrastructure for the query processing across distributed and heterogeneous data sources that may not conform to a single data model. Ontology-based semantic integration approaches are promising to achieve robust and flexible data integration framework [16][17]. Deficiency of secure methods for sharing vulnerable data may result in unwanted disclosure, security threats or irreversible losses [18]. Hence, powerful and secure data management framework is a crucial requirement to secure data while being able to query and retrieve RDF data. Security concerns in processing, storing, and transferring of confidential or sensitive data are a serious concern that could cause ethical, intellectual property and privacy issues [19][20]. Access control management is another important concern needs to pay attention while providing a secure environment to prevent data misuse or theft [21]. By providing proper authorization, we can ensure only specific individuals have access to data, making data unavailable to all others [22].

### 2.2. Addressing accuracy and quality of data issues

Today large volumes of data are growing exponentially in every domain of science, and valuable knowledge is driven by the capability of collecting and processing of this data. Although data is accumulated and analyzed very quickly in information systems, there is a need to establish approaches that can guarantee a certain degree of data quality [23][24]. Problems of data quality and trustworthiness must be addressed before preparing the data for subsequent operations. To tackle this challenge researchers must analyze the quality of data and assure that data is accurate, complete and consistent in order to decrease operational risk [25][26].

## 2.3. Addressing data operation and data access issues

The widespread heterogeneous data available in big data systems also have made security critically important. The gap between data processing and security requirements has given rise many issues and concerns. By compromising fragile defenses, attackers can strike the system not only from outside the system but also from within it [27]. An attacker may attempt to obtain unauthorized read/write permissions against the stored data objects or attempt to reveal or derive the credentials of account owners and try to perform an unauthorized read or write operation on stored data. From data operation perspective, there are four basic operations: write, read, delegate, and revoke [28]. The user needs to be provided with a valid access key to ensure a specific data is being accessed for read or write operation by that authorized user [29]. Also, for a robust auditing system, super-users have privileges to grant delegation and revocation to other users and also customize access rules for extracting queries [30]. Authorization is an even higher priority when private or sensitive data is stored in a multi-user environment, as insufficient authorization and access management system allows attackers to gain access to data and compromise the consistency of the system [31].

## 3. Study design and experiment setup

### 3.1. Ontology in plant science

The term Ontology is originated from philosophy where it refers to the nature of existence. Particularly, ontology was used to provide a semantic framework for representing knowledge using ontology representation languages. Currently, several ontology representation languages have been proposed including RDF, RDFS, and OWL to capture the semantics of the domain of study [32]. The bio-ontology emerged for enhancing the interoperability within biological knowledge with the best practices on ontology development [33].

The most cited bio-ontology is the Gene Ontology (GO) [34][35], which is a tool for the unification of biology developed by Gene Ontology Consortium in 2000, present more than 30000 species-independent control vocabularies for describing gene products including plants. Plant Ontology (PO) focused on developing and sharing unambiguous vocabularies for plant anatomy and morphology. PO consist of two sub-categories: the plant structure ontology and the growth and developmental stages ontology [36]. By defining classes of entities, logical relations, properties, constraints and range axioms, botany and plant science researchers are able to understand, share and reuse knowledge in a machine or computer interpretable content, enabling them to detect and reason biologically common concepts in heterogeneous datasets [37].

The Plant Science Ontology main goal is to design and develop a semantic framework in order to support computerized reasoning. With the help of ontologies, scientists are able to employ PO or GO as a general reference to semantically link large amounts of plant phenotype and genotype data together. However, knowledge engineering requires extensive knowledge of different domains such as biology, engineering and also standard ontology languages [38][39].

## 3.2. Tools and implementation of ontology

There is no specific method for modeling and building a domain ontology, and the majority of the best methodologies for developing an ontology depends on the purpose of research. For building an ontology, researchers should consider three features. First, identifying the domain and scope of the ontology. Second, choosing the language and logic to construct the ontology. Finally, identifying key concepts of resources (nouns), relationships (verbs), domain and range axioms.

We study Arabidopsis Thaliana as a reference plant for building our ontology. Arabidopsis Thaliana is the best investigated flowering plant species, belonging to the Brassicaceae family. It has been chosen as one of the most widely used model plant organism for studies in plant research in areas such as developmental and molecular genetics analysis, population genetics and genomics for many years. Its significant properties such as short regeneration time and simple growth requirements maks it desirable for model plant studies. Therefore, studying biological processes in this species is important for gaining information about plant science and for utilizing of this knowledge to other relevant plants species.

We have used the Protégé as the principal ontology authoring tool in our ontology-based application. Protégé is an IDE developed at Stanford University by Stanford Medical Informatics team. It is a free and open-source ontology platform that enables users to create and populate ontology and formal knowledge-based applications more straightforward. There exist many plug-ins for Protégé offering a number of powerful features. We used the OWLViz plug-in to visualize Protégé ontologies. OWLViz is a powerful and highly configurable extension providing a graphical representation of the semantic relationships for helping users to visualize classes in an OWL ontology. OWLViz creates a knowledge-based graph of the classes to different formats such as JPEG and PNG [40].

The process of building ontology for Arabidopsis Thaliana with the top-down approach is described in detail in this section. This ontology acts as a basis for researchers to conduct queries and reason in a knowledge-based environment. All OWL classes inherit from a single root class called owl:Thing. The class owl:Thing represents the concept of any user-defined class or individuals in order to facilitate reasoning, as illustrated in Figure 1.

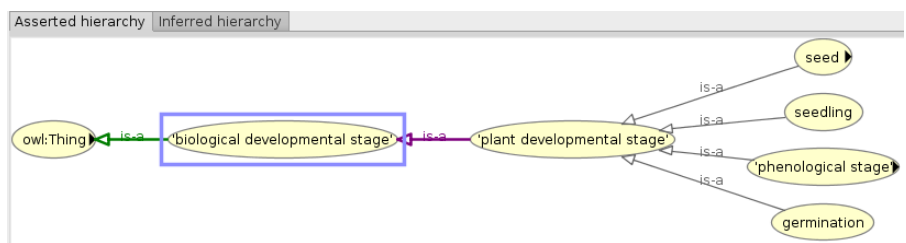

Figure 1: Class Tree of Arabidopsis Thaliana in Protege

Owl:Thing is developed by W3C located at http://www.w3.org/2002/07/owl as part of OWL vocabulary and is equivalent to rdfs:Resource. The most basic part of the ontology for Arabidopsis Thaliana ontology is super classes such as biological developmental stage, biological process, and biochemical process as shown in Figure 2. Each super class consists of sub classes which arranged in an inheritance hierarchy. The second level in ontology are subclasses which provide more refined and detailed information about superclass, such as germination, life span, seed and seedling.

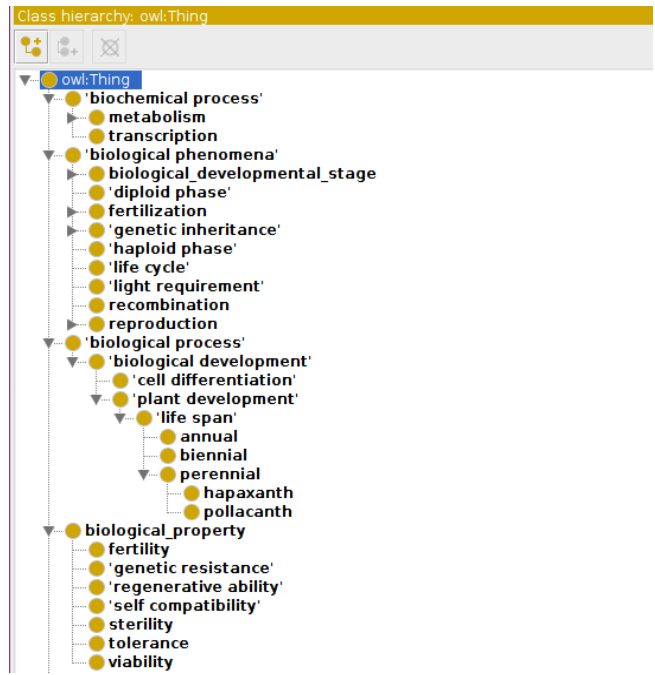

Figure 2: Calss Hierarchy of Arabidopsis Thaliana in Protege

OWL properties express association between two entities of a domain. Different types of properties are used to link between concepts. Figure 3 shows the list of declared properties of Arabidopsis Thaliana ontology in Protégé for this study. Some of the relations used in this ontology are <growsIn>, <hasPart>, and <hasVariant>.

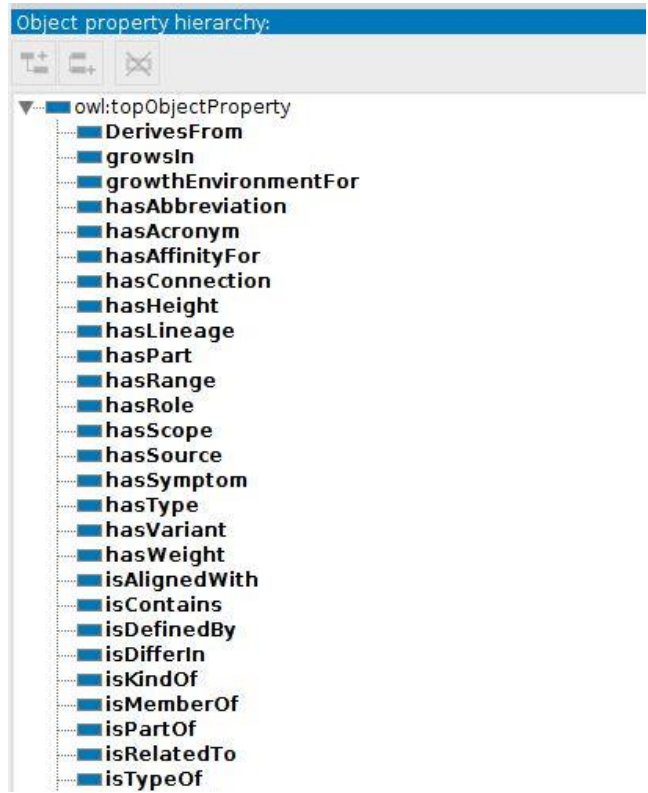

Figure 3: Object Properties of Arabidopsis Thaliana

Domain and range constraints of the properties aid to precisely describe a representation of knowledge. Domain indicates the type of individuals that can be the subject and range specifies the type of individuals that can be the object within the RDF triple. The individuals are the members or instances of a class with certain constraints as illustrated in Figure 4.

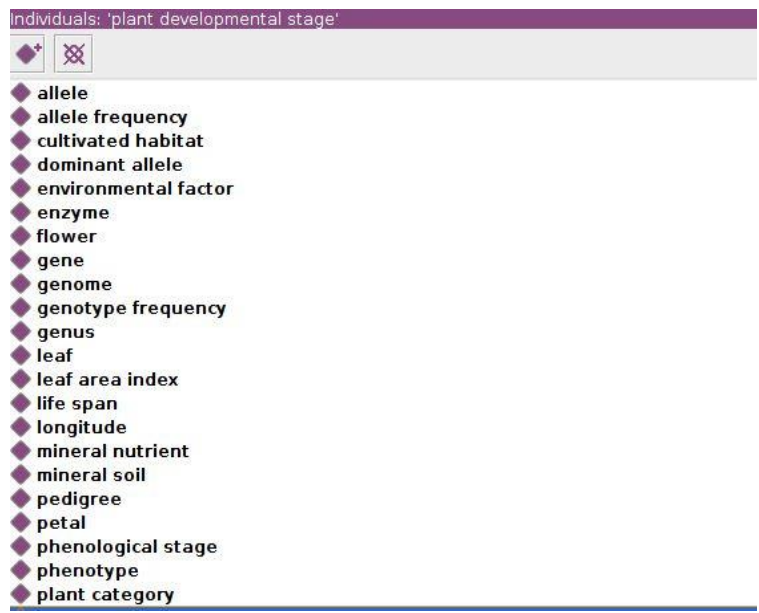

Figure 4: Individuals of Arabidopsis Thaliana Ontology

After preparing the ontology, we have used Apache Jena Fuseki server, a HTTP-based query engine, for executing the SPARQL queries over the web of linked data and extract triple information on the web page. Apache Jena Fuseki server is a SPARQL server providing hosts for persistent storage or in-memory storage of the datasets and aimed toward supporting developers with building practical semantic web applications. The class biological property, for example, has several subclasses. Some subclasses of biological property that are used to describe the class are genetic resistance, regenerative ability, seed compatibility, tolerance and viability. A wide variety of object properties are used to describe relationships between classes and subclasses. The object property growsIn describe the plant-specific requirement for growth and development. The object property maxHeight indicates the plant sample height growth pattern. The ontology of Arabidopsis Thaliana in this study provides detailed information pertaining to the morphology and developmental stages.

### 3.3. Ontology evaluation

After building the Ontology, we should assess the consistency and quality of the ontology using Protégé reasoner. Ontology evaluation and validation ensure the avoidance of excessiveness concepts, terminological ambiguity, incompatible subclass relationships. Ontologies need to be validated to confirm a standardized OWL profile, the expressivity of ontology, as well as the consistency of structure described in the model with expected semantics to support the exchange of information efficiently. Further consultation with domain experts is needed to examine the scope of the proposed ontology [41][42]. To evaluate the overall performance of the implementation and ontology syntax, we have used OntoCheck plugin to evaluate the consistency, conciseness, and correctness of ontology automatically. OntoCheck is an open source plug-in developed at the University of Freiburg, and it is currently one of the W3C OWL official validating tool [43] [44].

### 4. View layer on semantic data integration using distributed ledger technology

The idea of view-based integration systems presents a single point (the view layer) of a query and data access for a domain of a specific ontology. Users would execute large numbers of expensive queries over the unified view and get back results in order to get the best possible query performance. The main goal of using view layer is to satisfy user-specific requirements while processing data in a timely and secure manner through an integrated framework.

On the other hand, for preventing attackers from exploiting vulnerabilities, a newly-initiated approach is needed to verify user's authentication before providing access to the stored data and enhance security. Employing a private blockchain as a secure, decentralized and distributed corporate ledger system can ensure the correctness and completeness of data in the view layer [45][46]. The blockchain is a cryptographic technology that has a secure and resilient architecture and is used to track the records of data ownership. Users on a network create a transaction history.

The blocks of data subsequently join together to create chains of blocks, storing a transaction history among users without the need for a centralized control [47][48]. Once a transaction is published to the blockchain and confirmed as accurate, it cannot be reversed or destroyed and is immutable. We can benefit from this approach to reason about whether an individual is permitted to perform a read/write action.

Each block on the blockchain references to the previous one and contains information regarding the transactions and data about the block. Figure 5 illustrates a blockchain and contents of a block. Usually, the block header contains information about the version of the block and a hash of the data in the block [49].

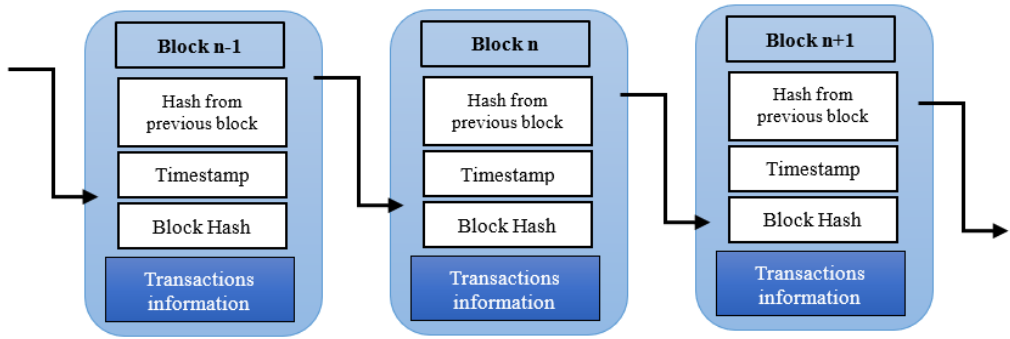

Figure 5: Structure of Blockchain showing chained blocks

The significant advantage of blockchain is smart contracts. Smart contracts as a fundamental mechanism in developing access control policies allow users to control access of data by writing a pre-defined set of criteria and condition [50][51]. The smart contracts will automatically execute the terms of the contract. These contracts increase efficiency by being able to interact with other contracts, storing data, and making decisions. Smart contracts' information is distributed throughout the blockchain for the cost of a small token and automatically executed upon occurrence of the event [52]. The view layer is transparent to the end users and is able to detect compromised or malicious users; thus, it provides a framework that is able to solve security concerns such as identity management, access control, trust, and authorization. Furthermore, this approach does not allow the user's authentication to be forged or modified and hence remains highly resilient against intrusion [53] [54].

### 4.1. The architecture of the view layer

The idea of our representation model for storing and retrieving views of queries calls for a layer of abstraction over the RDF data sources [55]. This layer of abstraction is called a 'view layer' as illustrated in Figure 6. View layer is dedicated for managing views including storing and retrieving views of queries and also keeping the views updated from the data sources in order to support real-time semantic query processing of the ontology. The views are generated and maintained at the back-end, and each view may be used by several users as a shared view.

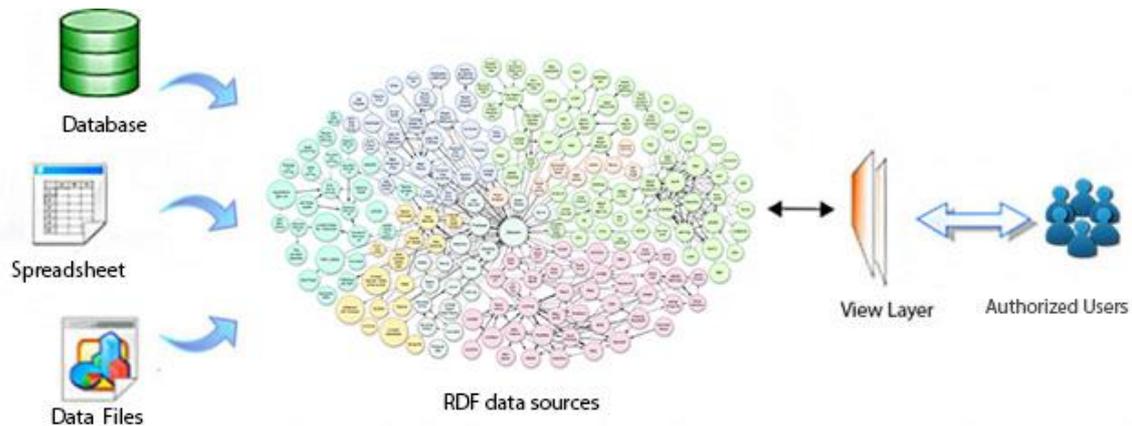

Figure 6: Data integration system and view layer

View layer also prevents improper access of unauthorized users to sensitive or confidential information based on users corresponding credential levels. In view layer architecture, users only allowed to have access to data by querying a view of the data, they do not have permission to directly query or access over the original data sources. In this scenario, users with different level of authorization are allowed to have access the same queries, but not the same results in order to prevent information leakage and security breach.

Figure 7 illustrates an overview to view layer architecture: data flows from a user that sends a query to the view layer. The users' request will be sent to a module called a blockchain access control. The blockchain access control module will examine the users' privileges and make a decision on what kind of access should be granted to users' request. After verifying the level of permission and authorization role of the user, the coordinator module translates or reformulates queries into equivalent queries with respect to the data integration schema and user credential. Query processor creates the query result in a database for views in the view manager module and then return the answers to the users while satisfying all security requirements. When a user sends a read or write request to have access to the RDF data store, based on system smart contract rules, the access level will be analyzed and if user is eligible to have access to data, the requested query will be executed. When a request to the database is accepted, user information and transaction information will be added to the block in order to keeps track of all transactions in a distributed ledger.

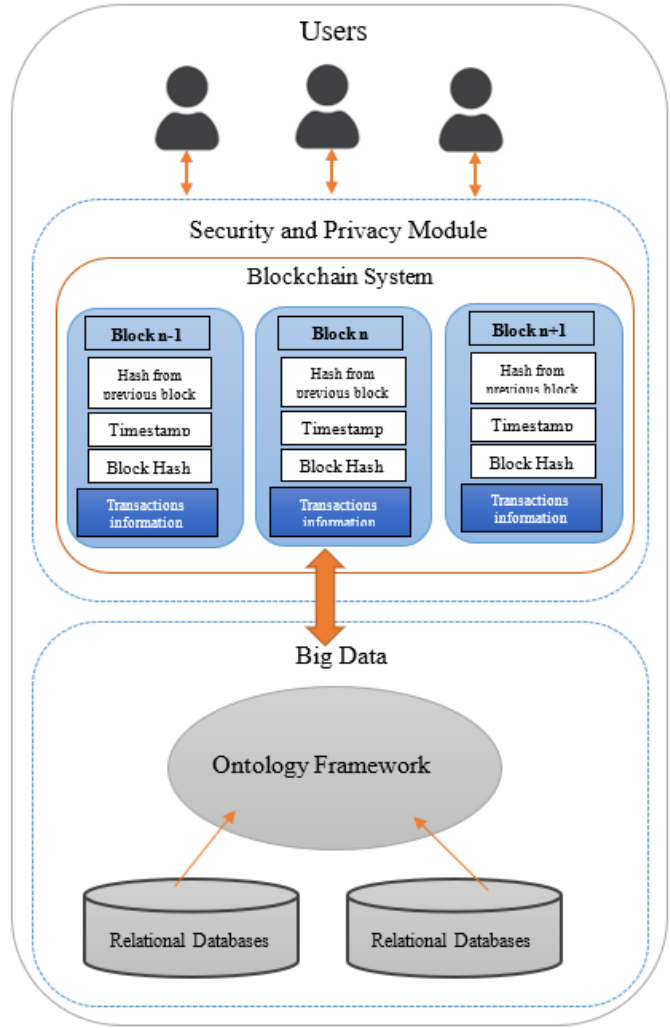

Figure7: An overview of View Layer Architecture

Blockchain model in the proposed semantic data integration architecture avoids compromising unauthorize data access and protects transactional privacy [22] using a Role-Based Access Control (RBAC) approach. In RBAC, roles are assigned to users and privileges are assigned to roles in order to determine if a user is eligible to have access to data [26] [27]. As demonstrated in Figure 8, a participant in the system based on the type of role is allowed to have access to a specific data. Roles is issued by the appropriate authority in order to determine whether or not a user (or group of users working together) should be trusted to have access to a specific data. The role-based access control using smart contracts on data verifies the access role of a user. In this architecture, when a user sends requests to have access to data, a rule in smart contract is triggered to check the user assigned role to that data. If the user has a role that has access privileges then, the user is allowed to view the data.

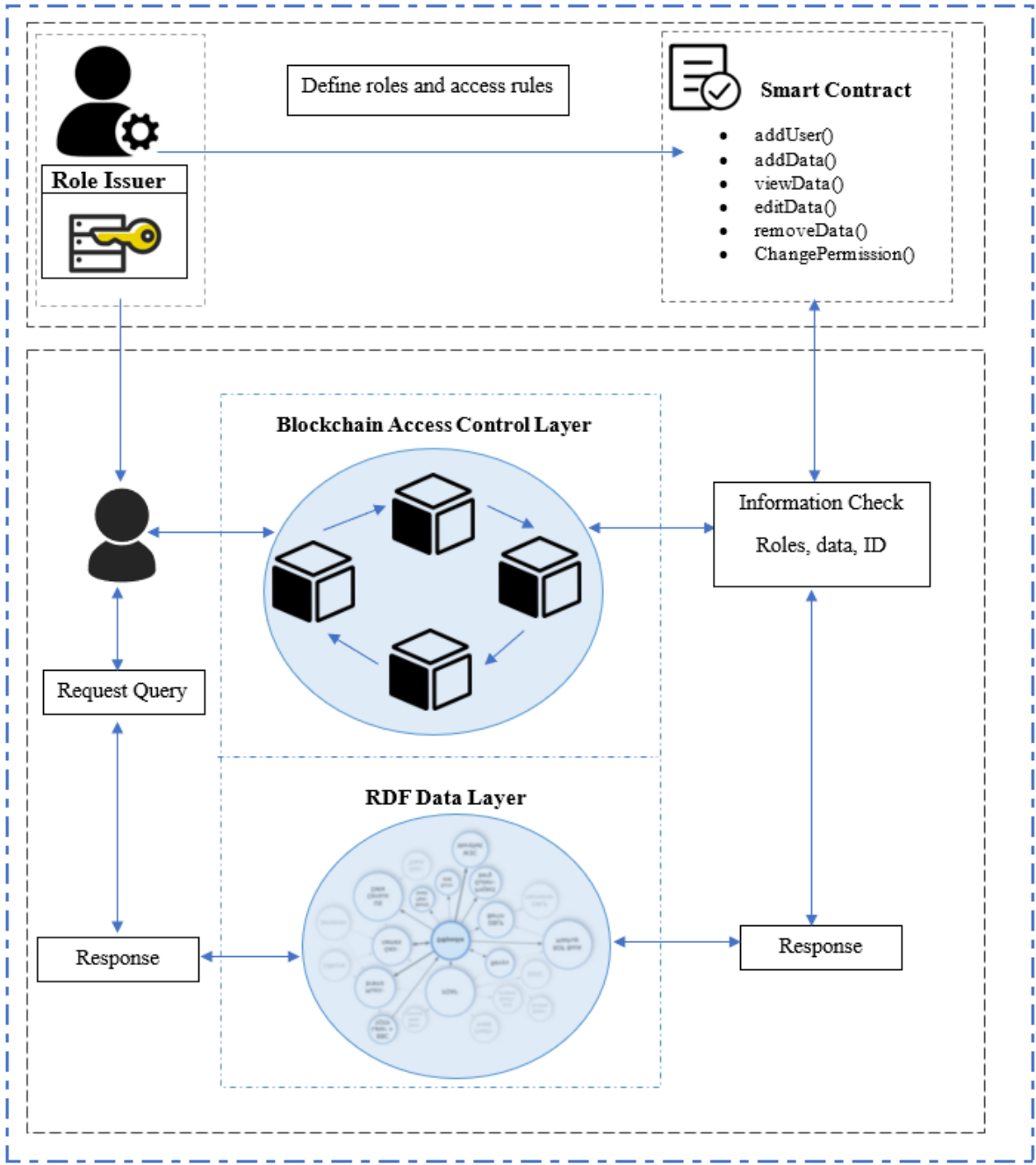

Figure 8: Blockchain Role-Based Access Control (BRBAC) system

In the Blockchain Role-Based Access Control (BRBAC) system, the smart contract file contains rules for data and transactions including request, grant, revoke, verify and also view. The assumption is that access privileges based on defined roles are granted to system users by role issuer to a subset of the RDF data [56]. When a user attempts to access a data, smart contract verifies that the user has a role with appropriate access rights to that data and based on the response from smart contract, the user is allowed or denied having access to a subset or view of data. Using Blockchain Role-Based Access Control approach, a system administrator can assign IDs to data and enable blockchain to control access between users and data store by verifying user data access level using assigned ID. The IDs can be defined to represent a specific data or a query that allows access to a view of data. The blockchain access control module is separated from the data store and hence, can be used as an access control layer to the existing infrastructure.

## 5. Conclusion

Large volumes of data are generated by a wide variety of sources. However, security challenges and complications in processing, storing, and transferring of confidential or sensitive data exist that need to be addressed. These challenges could cause ethical, intellectual property and privacy issues. There is therefore a need for efficient secure privacy preserving systems. In this study, we implemented a semantic data integration framework with a case study of plant ontology to ensure data quality and trustworthiness in extracting and processing queries over the RDF triples stores. Also, we proposed a secure view-based layer architecture using distributed ledger technology to prevent unauthorized read/write permissions against stored data. The goal of this architecture is to provide a secure and private method in allowing access to data to users with specific privileges by integrating blockchain technology with existing semantic data integration framework.